\documentclass[floats,floatfix,showpacs,amssymb,prd,twocolumn,superscriptaddress,nofootinbib,nolongbibliography,reprint]{revtex4-2}

\usepackage{amssymb,amsmath,verbatim,mathtools,needspace,enumitem,etoolbox,graphicx,physics,microtype,afterpage,xspace,tabularx,lmodern,multirow}
\usepackage{gensymb}
\usepackage[dvipsnames, usenames]{xcolor}
\definecolor{linkcolor}{rgb}{0.0,0.3,0.5}
\usepackage[unicode, colorlinks=true, linkcolor=linkcolor, citecolor=linkcolor, filecolor=linkcolor, urlcolor=linkcolor, linktocpage, breaklinks]{hyperref}
\usepackage[all]{hypcap}
\usepackage[T1]{fontenc}
\usepackage[utf8]{inputenc}
\usepackage[usenames,dvipsnames]{xcolor}
\hypersetup{colorlinks=true,citecolor=romared,linkcolor=romared,urlcolor=romared}

\definecolor{romared}{RGB}{142,0,28}

\newcommand{\be}{\begin{equation}}
	\newcommand{\ee}{\end{equation}}

\def\be{\begin{equation}}
	\def\ee{\end{equation}}
\newcommand{\beq}{\begin{eqnarray}}
	\newcommand{\eeq}{\end{eqnarray}}

\usepackage{aas_macros}
\usepackage{makecell}
\usepackage{soul}
\usepackage{blindtext}
\usepackage{hyperref}

\newcolumntype{Y}{>{\centering\arraybackslash}X}

\begin{document}

	\title{Mergers of Maximally Charged Primordial Black Holes}
	
	\author{Konstantinos Kritos}
	\email{kkritos1@jhu.edu}
	\affiliation{Department of Physics $\&$ Astronomy, The Johns Hopkins University, Baltimore, MD 21218, USA}
	\author{Joseph Silk}
	\email{silk@iap.fr}
	\email{jsilk@jhu.edu}
	\email{joseph.silk@physics.ox.ac.uk}
	\affiliation{Institut d’Astrophysique de Paris, UMR 7095 CNRS $\&$ UPMC, Sorbonne Universit$\acute{e}$, F-75014 Paris, France}
	\affiliation{Department of Physics $\&$ Astronomy, The Johns Hopkins University, Baltimore, MD 21218, USA}
	\affiliation{Beecroft Institute for Particle Astrophysics and Cosmology, Department of Physics, University of Oxford, Oxford OX1 3RH, UK}
	
	\date{\today}
	
	\begin{abstract}
	
	Near-extremal primordial black holes stable over cosmological timescales may constitute a significant fraction of the dark matter. Due to their charge the coalescence rate of such black holes is enhanced inside  clusters and the non-extremal merger remnants are prone to Hawking evaporation. We demonstrate that if these clusters of near-extremal holes contain a sufficient number of members to survive up to low redshift, the
	hard photons from continued evaporation begin to dominate the high energy diffuse background. We find that the diffuse  photon flux can be observed for a monochromatic mass spectrum of holes lighter than about $10^{12}\rm g$.
	We place upper bounds on their abundance respecting the current bounds set by gamma ray telescopes. Furthermore, the gravitational wave background induced at the epoch of primordial black hole formation may be detectable by future planned and proposed ground-based and space-borne gravitational wave observatories operating in the mHz to kHz frequency range and can be an important tool for studying light charged primordial black holes over masses in the range $\rm 10^{12}g - 10^{19}g$.
	
	\end{abstract}
	
	\maketitle
	
	\section{Introduction}
	
	The nature of dark matter (DM) is an outstanding mystery in  modern cosmology. One possible explanation is that DM is composed of massive compact halo objects, and black holes serve as one of the candidates. These are assumed to form in the early Universe, and are more appropriately termed primordial black holes (PBHs). According to the standard scenario, PBHs form during the 
	radiation-dominated era by the gravitational collapse of large over-densities in the density field shortly after the Big Bang and have been hypothesized to account for the DM in the Universe \cite{1974MNRAS.168..399C,1975Natur.253..251C,1974A&A....37..225M,1975ApJ...201....1C,Clesse:2015wea}.
	Interest in primordial black holes was revived after the first gravitational wave events observed by the LIGO and Virgo collaborations \cite{LIGOScientific:2016aoc} were found to involve higher mass black holes than routinely observed in X-ray binaries.
	Data analysis of these gravitational waves indicate that they correspond to the coalescence of stellar-mass black hole binaries. Some or even all of them could be of primordial origin and might explain the merger rate of the detected events \cite{Bird:2016dcv,Clesse:2016vqa,Ali-Haimoud:2017rtz,Raidal:2017mfl,Jedamzik:2020ypm,Wong:2020yig,Franciolini:2021tla}. 
	
	Black holes are thought to be the simplest objects in the Universe in the sense that they can be fully described in terms of only three parameters: mass, spin angular momentum and electric charge, a proposition that is often referred to  as the {\it no-hair theorem}. The most general geometry describing such a black hole is the Kerr-Newman metric which reduces to the Reissner-Nordstr{\"o}m metric for zero spin and the Kerr metric for a neutral rotating black hole. So far, most interest in PBHs has been devoted to the case of uncharged black holes. Nevertheless, in principle black holes are physically allowed to carry an electric charge \cite{Zajacek:2019kla}, even though stellar mass black holes formed via some astrophysical mechanism such as from core collapse of massive stars are presumed to be electrically neutral. Depending on the initial conditions in the early Universe and the density perturbations that gave rise to primordial black holes, it is possible that PBHs formed with some residual electric charge. Since the total charge in the density field averaged over some comoving volume is expected to be zero, there is correspondingly an equal number of positively and negatively charged primordial black holes per unit comoving volume. More generally, depending on the power spectrum of curvature perturbations that is responsible for PBH formation, PBHs may have any mass within a range spanning more than twenty orders of magnitude. 
	
	The abundance of PBHs is constrained by various observations, for instance gravitational microlensing, cosmic microwave background radiation and dynamical considerations, and it is generally concluded that PBHs do not comprise the totality of DM in certain mass intervals \cite{Carr:2020xqk}.	
	For light neutral PBHs with masses below $\sim10^{17}\rm g$, the most stringent constraints come from evaporation arguments \cite{1976ApJ...206....1P,1991Natur.353..807H,Carr:2009jm,Arbey:2019vqx,Ballesteros:2019exr}. Black holes evaporate as thermal blackbodies \cite{Hawking:1975vcx} and those with a mass smaller than $\simeq5\times10^{14}\rm g$ are expected to have evaporated on a timescale smaller than the age of the Universe \cite{1976PhRvD..13..198P}. If primordial black holes are cosmologically stable, then they can avoid evaporation and such a constraint can be evaded. 
	
	An extremal black hole holds the largest possible charge, has a vanishing Hawking temperature and does not evaporate. But extremal black holes promptly deviate from the extremality condition due to Schwinger discharge \cite{1975CMaPh..44..245G} which is a form of athermal emission by spontaneous electron-positron pair production driven by the large electric field near the event horizon. To consider extremal black holes that are stable against Schwinger discharge, we assume the black holes to be charged under dark electromagnetism with a heavy dark electron as suggested in Ref.~\cite{Bai:2019zcd}. In particular, the minimum mass of an extremal black hole that survives for a Hubble time can be very light provided that the mass of the dark electron $M_{e'}$ is correspondingly large. Hence, according to the previous Ref. as long as  $M_{e'}c^2>2.6\times10^{9}\mathrm{GeV}\cdot(10^{15}{\rm g}/m)^{1/2},$ then the black hole of mass $m$ is stable over the age of the Universe.
	Nevertheless, mergers of these objects give rise to non-extremal black holes which have a non-zero surface gravity and thus evaporate producing highly energetic photons, therefore some constraint on their abundance is expected. Since not all of the extremal black holes merge to supply evaporating objects, the constraint is presumed to be stringent for very light PBHs.
	
	Mergers of charged black hole binaries have not been considered extensively in the literature. Head-on collisions of charged black holes were studied in Refs.~\cite{Zilhao:2012gp,Zilhao:2013nda} and a recent numerical relativity calculation of a binary composed of charged black holes has been performed in Ref.~\cite{Bozzola:2021elc} for non-extremal black holes with charge-to-mass ratio less than 0.3. The charge-to-mass ratio of stellar-mass black holes charged under dark QED was also bounded above relying on the LIGO-Virgo events \cite{Bozzola:2020mjx,Cardoso:2016olt}. Moreover, Ref.~\cite{Liu:2020cds} calculated the evolution of charged binaries in the early inspiral phase within the Newtonian framework and showed that the merger rate from charged primordial black holes is enhanced. We utilize the coalescence time-scale approximation formulae from this paper and demonstrate an enhanced late-time merger rate  for dynamical assembly within clusters.
	
	We assess the possibility that maximally charged primordial black holes which are stable over the history of the Universe may constitute a significant fraction of the dark matter. Potential signals associated with their formation and late time evolution are evaluated. We are agnostic on the formation of these extremal black holes and test the possible signatures such a population would create if it were present.
	For simplicity, we take a monochromatic mass function. Clustering of these charged holes after the epoch of recombination enables the dynamical formation of black hole binaries and associated capture events that are a source of mergers. 
	The two-body capture process between oppositely charged objects or electromagnetic bremsstrahlung is the primary mechanism we consider that leads to binary creation. We further show that the merger rate is enhanced due to the charges of the black holes and the relaxation timescale of these clusters is smaller than that of an equivalent cluster composed of neutral compact objects. The remnants of these mergers will have a larger mass, and since their progenitors are of opposite charge, they tend to neutralize, so the remnants are non-extremal. As a consequence, it is expected that they will evaporate by emitting Hawking radiation and produce a signal potentially observable by gamma ray telescopes. We employ the publicly available {\tt BlackHawk} software \cite{Arbey:2019mbc} to calculate photon emission spectra from Hawking evaporation. Even though we take the extremal primordial black holes to form with no spin as is suggestive in recent literature \cite{Mirbabayi:2019uph,DeLuca:2019buf}, merger remnants  still inherit the orbital angular momentum of the progenitor binary which leaves an imprint on the emission spectrum. Another detectable signal that we calculate is the gravitational wave background induced at the epoch of primordial black hole formation in the early Universe. Such a background could be detectable by current and future proposed or planned gravitational wave observatories.
	
	In the following sections, we will develop and implement the physical ideas. In Section~\ref{Sec:setup}, we calculate the merger rate in primordial black hole clusters and demonstrate that it is enhanced for charged black holes. This merger rate plays the role of the source for the production rate of evaporating black hole remnants which contribute to the diffuse gamma ray background. The gamma ray flux and induced gravitational wave background are presented in Sec.~\ref{Sec:signals} as well as constraints on the abundance of such maximally charged holes. Finally, we summarize and give our conclusions in Sec.~\ref{Sec:conclusions}.
	
	\section{Merger rate of extremal black holes}
	\label{Sec:setup}
	
	Assuming a Gaussian density field in the early Universe, it has been demonstrated in Ref.~\cite{Ali-Haimoud:2018dau} that the spatial distribution of the PBHs is random and follows a Poisson distribution. 
	PBHs do not form in clusters and only Poisson cluster in the late Universe when perturbations start to grow after the epoch of recombination \cite{Carr:2018rid,Inman:2019wvr}.
	The clusters eventually virialize, as in the noncharged case \cite{Jedamzik:2020omx} and form a neutral environment of light charged PBHs. If we denote by $m$  the mass of the cluster members and $N$ the number of charged black holes in the cluster, then according to the Virial theorem, we can relate the velocity dispersion $\sigma$ with the previously mentioned parameters, $\sigma^2=(6/5)GNm/R$\footnote{The electromagnetic potential energy is neglected as it is subdominant to the gravitational potential due to screening effects.}. Here, the symbol $R$ denotes the size of the system. 
	As with clusters of neutral black holes, these are expected to evaporate with time and the cluster has a finite lifetime. In Appendix~\ref{App:relax} we prove that the lifetime of such a cluster depends on the number of members it contains and is given by Eq.~\eqref{Eq:ClusterLife}. Therefore, such a cluster survives for more than a Hubble time if it hosts at least $N\simeq2000$ members (see App.~\ref{App:relax}). As we will show in Sec.~\ref{subSec:signals_Hawk} the lifetime of a cluster plays an important role in determining whether the diffuse gamma ray background produced by the evaporating merger remnants could be detectable.
	
	For simplicity, we take all black holes to be initially isolated inside the clusters. The inclusion of proto-binaries only enhances the merger rate and our assumptions should be conservative and moreover a fraction of these would not contribute to the merger rate since binary-single encounters would disrupt them \cite{Jedamzik:2020ypm,Raidal:2018bbj,Vaskonen:2019jpv}. As a consequence of two-body encounters, similar to Rutherford scattering, if two oppositely charged black holes pass each other with a small separation,  capture can result in the formation of  a compact binary. The dissipative mechanism is electromagnetic in nature and associated with the acceleration of charges along a hyperbolic encounter. Simultaneously, such an unbound two-body system also generates gravitational waves due to an accelerating mass quadrupole moment. We calculate the total electromagnetic radiation $\delta E_{em}$ emitted during such an interaction in App.~\ref{App:bremss}, given by Eq.~\eqref{Eq:EnergyEM} within the dipole radiation approximation. The total energy emitted in gravitational waves $\delta E_{gw}$ during an unbound orbit was calculated in Ref.~\cite{1977ApJ...216..610T} (see Eq.~\eqref{Eq:EnergyGW} and also Ref.~\cite{1978ApJ...220.1107T}). The corresponding 
	cross-section for gravitational captures is well known in the literature as a potential mechanism through which black hole binaries are assembled \cite{1989ApJ...343..725Q,Mouri:2002mc}.
	
	To estimate the merger rate due to bremsstrahlung events, we need to calculate the cross-section for capture. The condition for two oppositely charged black holes to be captured is that the total energy released, the sum of electromagnetic and gravitational bremsstahlung $\delta E=\delta E_{em}+\delta E_{gw}$, be larger than the energy of the two body system at infinity, which is the kinetic energy of the reduced body $T=m\sigma^2/2$. The relative velocity between two members of the cluster is $\sqrt{2}\sigma$. This condition translates into a formula for the maximum allowed pericenter of interaction for the capture to occur. A detailed calculation of this pericenter radius in the gravitational ($\delta E_{gw}\gg \delta E_{em}$) and electromagnetic ($\delta E_{gw}\ll\delta E_{em}$) regimes respectively gives
	\begin{align}
		r_{p_\mathrm{max}}/r_H = 
		\begin{cases}
			4.9\cdot\left({c\over\sigma}\right)^{4/7}&\ \mathrm{gw\ regime}\\
			2.6\cdot\left({c\over\sigma}\right)^{4/5}&\ \mathrm{em\ regime}
		\end{cases}.
		\label{Eq:pericenter}
	\end{align}
	We have defined the symbol $r_H\equiv Gm/c^2$ to be the event horizon radius of an extremal black hole with mass $m$. Typically, electromagnetic captures are expected to dominate over gravitational captures, since the loss cone for an em capture is effectively larger. In comparison to the mean separation $R/N^{1/3}$ between members of the cluster, we remark that for non-relativistic environments for which $\sigma\ll c$ the pericenter for capture is a much smaller quantity. Consequently, for capture events between black holes we can regard the close interaction occurring within the focusing regime and neglect the geometric part of the cross-section.
	
	Given the maximum allowed pericenter for efficient bremsstrahlung (Eq.~\eqref{Eq:pericenter}), the 
	cross-section for capture when the focusing term dominates in our case reduces to\footnote{The 
	cross-section for a process $X$, given that it occurs when the maximum allowed pericenter of interaction is $r_{p_\mathrm{max}}$ for this process, is given by the following formula 
	\begin{align*}
		\Sigma_{X}=\pi r_{p_\mathrm{max}}^2\left(1+{2\alpha/\mu\over r_{p_\mathrm{max}}v_\mathrm{rel}^2}\right)
		\label{Eq.CrossSection}
	\end{align*}
	as the sum of the geometrical ($\propto r_{p_\mathrm{max}}^2$) and focusing ($\propto r_{p_\mathrm{max}}$) terms. The symbol $v_\mathrm{rel}$ is the relative velocity between the two interacting species. Further, the interaction potential is $U=-\alpha/r$ and $\mu$ is the reduced mass. For future reference, the velocity-averaged cross-section in the focusing dominated regime is $\langle\Sigma_{X}v_\mathrm{rel}\rangle=2\sqrt{3\pi}\alpha r_{p_\mathrm{max}}/(\mu\sigma)$. The angular brackets $\langle...\rangle$ denote the average performed over the relative velocity $v_\mathrm{rel}$ taking the Maxwell-Boltzmann distribution.} \ $\Sigma\simeq4\pi Gmr_{p_\mathrm{max}}/\sigma^2$. 
	As we show in App.~\ref{App:CaptureMeansMerger}, capture events lead to the prompt merger of the compact black hole pair since it tends to form highly eccentric binaries and energy dissipation is very efficient. So, binaries formed via capture coalesce before they get disrupted by the interaction with a third object. Therefore, calculating the capture rate gives us an estimate for the merger rate. Since we are in the extremal case, only oppositely charged black holes may form a bound pair (see Eq.~\eqref{Eq.PairEnergy} and the discussion following Eq.~\eqref{Eq.mergerTime}) and the capture rate density is given by $d\Gamma_\mathrm{cap}/d^3r=n_+n_-\langle\Sigma_\mathrm{cap}v_\mathrm{rel}\rangle$. The symbols $n_{\pm}$ denote the number densities of positively and negatively charged black holes. Since the cluster is overall electrically neutral, we take these to be equal, $n_+=n_-=3N/(8\pi R^3)$. Taking a uniform mass profile\footnote{We have performed the same calculations under the assumption of a Plummer model for the mass density \cite{1911MNRAS..71..460P}. However the resulting merger rate differs from the one we show here only by about $20\%$ and the exact profile has a small effect on our results.} direct evaluation of the merger rates assuming electromagnetic and gravitational captures gives
	\begin{align}
		\Gamma_\mathrm{cap}\tau_\mathrm{cross}&\simeq{r_{p_\mathrm{max}}\over r_H}\left({\sigma\over c}\right)^2\nonumber\\&=
		\begin{cases}
			4.9\cdot\left({\sigma\over c}\right)^{10/7}&\ \mathrm{gw\ regime}\\
			2.6\cdot\left({\sigma\over c}\right)^{6/5}&\ \mathrm{em\ regime}
		\end{cases}
	\end{align}
	where $\tau_\mathrm{cross}=R/\sigma$ is the crossing time in the cluster. Utilizing Eq.~(4) for the velocity dispersion from Ref.~\cite{Jedamzik:2020ypm} we find that
	\begin{align}
		{\Gamma_\mathrm{cap,\ em}\over\Gamma_\mathrm{cap,\ gw}}\simeq 0.5\cdot\left({\sigma\over c}\right)^{-{8\over35}}\simeq601\cdot N^{-{2\over105}}\left({m\over10^{10}g}\right)^{-{8\over105}}.
	\label{Eq:emVSgw}
	\end{align}
	That is, the ratio of the electromagnetic to the gravitational capture rate depends very weakly on the number of members and the mass scale of the black holes in the cluster, with a value of order a few hundred.
	This shows that electromagnetic bremsstrahlung is expected to dominate over gravitational bremsstrahlung and therefore we can neglect gravitational captures in the following analysis.
	
	To conclude this section, we evaluate the merger rate density per unit comoving volume. For this, we need to multiply the merger rate per cluster $\Gamma_\mathrm{mrg}\simeq\Gamma_\mathrm{cap,\ em}$ by the  number density $n_{cl}$ of cluster environments in the Universe. This number density can be related to other quantities which we can control better, parameters such as the fraction $f_\mathrm{ePBH}$ of the DM that is composed of extremal PBHs with mass $m$ and the number of members $N$ in each cluster. In particular, we have $n_{cl}=\rho_\mathrm{ePBH}/(Nm)$ where the mass density of extremal PBHs is parametrized in terms of the dark matter density parameter at the present epoch $\Omega_{DM}$ and the critical density $\rho_{cr}= 3H_0^2/(8\pi G)$ to close the Universe, $\rho_\mathrm{ePBH}=f_\mathrm{ePBH}\Omega_{DM}\rho_{cr}$. The symbol $G$ is the gravitational constant and $H_0=100h$ km/s/Mpc the Hubble constant at the present epoch. For the sake of simplicity, we take all clusters to have identical properties and be uniformly distributed in space. Cosmological parameters we use are motivated by the {\it Planck 2018} results \cite{Planck:2018vyg}. Provided with the previous definitions, the comoving merger rate density then becomes
	\begin{widetext}
	\begin{align}
		\gamma_\mathrm{mrg}&\simeq{630\over\mathrm{yr}\cdot\mathrm{m}^3}\cdot f_\mathrm{ePBH}N^{-2}\left({m\over10^{10}{\rm g}}\right)^{-2}\left({\sigma\over c}\right)^{21\over5}\simeq{48\over\mathrm{Myr}\cdot\mathrm{pc}^3}\cdot f_\mathrm{ePBH}N^{-{33\over20}}\left({m\over{10^{10}}{\rm g}}\right)^{-{3\over5}}.
		\label{Eq.MergerRateDensity}
	\end{align}
	\end{widetext}
	The second approximation above uses Eq.~(4) from Ref.~\cite{Jedamzik:2020ypm}. Notable is the dependence of the merger rate density on the mass of the PBHs. In particular, since $\gamma_\mathrm{mrg}\propto m^{-3/5}$, the smaller the mass scale we consider, the larger the merger rate density even if we keep constant the number of members per cluster and the abundance of PBHs. This is a consequence of the fact that as $m$ is reduced the number density of clusters increases to compensate for the decrease in cluster mass and keep the PBH density fixed relative to the DM.

	\section{Prospective signals}
	\label{Sec:signals}
	
	Mergers of cosmologically stable extremal black holes give rise to non-extremal black holes. These have a non-vanishing Hawking temperature and therefore evaporate. Over cosmic history, the merger rate supplies a source of evaporating objects which give rise to a diffuse extragalactic background of high energy photons.
	In this section, we calculate this flux as well as the gravitational wave background induced from the era of PBH formation as an alternative signal.
	\\
	\subsection{Gamma rays from the evaporation of merger remnants}
	\label{subSec:signals_Hawk}
	
	Electrically neutral black holes evaporate on a time-scale that depends on their mass scale. In the standard PBH scenario, the light PBHs are neutral and promptly start to evaporate after their formation. Constraints on their abundance can then be placed based on this assumption in order to respect observations of the diffuse extragalactic gamma ray background. In this work, however, since we have made the assumption that extremal PBHs are cosmologically stable against Hawking emission, it is the merger remnants that begin to evaporate after the progenitor binary has coalesced to produce them. As such, the production of evaporating holes persists over cosmic history and is ongoing for as long as extremal PBHs participate in clusters in order for the merger rate to be significant. Creation of evaporating holes ceases after clusters dissolve. The evaporating black holes then play the role of the source of gamma rays produced throughout the Universe.
	
	To calculate the cumulative flux of photons produced over cosmic history, we need to solve the following Boltzmann differential equations for the time evolution of the number density of sources $n_\mathrm{sources}$, here the merger remnants and the emitted particles $n_\mathrm{particles}$, in this case photons, in an expanding Friedmann-Robertson-Walker cosmological Universe:
	\begin{widetext}
	\begin{subequations}
		\begin{align}
			\dot{n}_\mathrm{sources}(t)+3H(t)n_\mathrm{sources}(t)&=\gamma_\mathrm{mrg}\Theta(t-t_\mathrm{begin})\Theta(t_\mathrm{end}-t)-n_\mathrm{sources}(t)\tau_\mathrm{Hawk}^{-1},\label{Eq.Boltzmann_a}\\
			\dot{n}_\mathrm{particles}(t;E)+3H(t)n_\mathrm{particles}(t;E)&=E{\cal D}(E)n_\mathrm{sources}(t).\label{Eq:nParticles}
		\end{align}
	\label{Eq.Boltzmann}
	\end{subequations}
	\end{widetext}
	In Eq.~\eqref{Eq.Boltzmann_a}, the symbol $\Theta$ denotes the Heaviside function. Moreover, the symbol $H(t)$ denotes the Hubble scale and the time moments $t_\mathrm{begin}$ and $t_\mathrm{end}$ represent the epoch of cluster formation and destruction respectively with the lifetime of the cluster being the difference $\tau_\mathrm{life}=t_\mathrm{end}-t_\mathrm{begin}$.
	Initial conditions to the above differential system are naturally given by $n_\mathrm{sources}(t=t_\mathrm{begin})=n_\mathrm{particles}(t=t_\mathrm{begin};E)=0$.
	
	During Hawking radiation, particles are emitted with different energies $E$ and form an energy spectrum. Hence the energy dependence of the particle number density on energy above is explicitly written $n_\mathrm{particles}(t;E)$.
	For the evolution of the number density of sources, there is one source term which is the merger rate assumed to be constant in redshift over the lifetime of the cluster. There is also a depletion term associated with the termination of the Hawking emission, and either the complete evaporation of the hole or the formation of a Planck-mass relic \cite{1987PhLB..191...51A}. The Hawking time-scale $\tau_\mathrm{Hawk}$ is calculated from Eq.~(12) of Ref.~\cite{1991PhRvD..44..376M}. Particles are produced and emitted for as long as there are sources, hence the form dependence of the source term in the right-hand-side of Eq.~\eqref{Eq:nParticles}. The particle emission energy spectrum ${\cal D}(E)=d^2{N}/(dEdt)$ is calculated implementing the {\tt BlackHawk} numerical package, which is publicly available \cite{Arbey:2019mbc}. When utilizing this package to calculate the emission spectrum from a single evaporating hole, we fix its mass and spin. 
	
	The hypothesis for the spin of the extremal black holes is that it is presumed small at the percent level \cite{Mirbabayi:2019uph,DeLuca:2019buf} and in this work we take it to be zero. Currently, numerical relativity simulations of oppositely charged black hole mergers have been performed. According to Ref.~\cite{Bozzola:2021elc} the difference in mass and spin of the remnant object increases with the charge-to-mass ratio relative to the uncharged case. The deviation in the remnant spin is only a few percent for charge-to-mass ratio $\le0.3$ and here we make the assumption that the merger remnant is a Kerr black hole with spin parameter $\chi_{\rm rem}\simeq0.7$. This choice is motivated by numerical relativity simulations of neutral equal-mass non-spinning black hole mergers \cite{Pretorius:2005gq,Hofmann:2016yih}. The mass of the remnant however seems to decrease significantly (see Fig.~14 of Ref.~\cite{Bozzola:2021elc}). However, we assume that the mass scale of the remnant would still be $m_{\rm rem}\sim m$.
	We stress again that these are reasonable assumptions, and a more detailed calculation in this context is beyond the scope of this work.
	
	For photon emission from Hawking radiation, there are two components to the energy spectrum. The primary component is associated with photons directly emitted by the black hole. The secondary component accounts for photons being created by the subsequent hadronization of quarks and gluon jets and the decays of other unstable particles being emitted. Therefore, ${\cal D}(E)={\cal D}_\mathrm{primary}(E)+{\cal D}_\mathrm{secondary}(E)$. The secondary component becomes significant and shows up only when the black hole is light and hot enough for it to produce the heavy particles of the Standard Model of particle physics. Here, we calculate both primary and secondary components considering only Standard Model degrees of freedom and take into account the instantaneous emission spectra.
		
	The diffuse particle flux observed today in units of cm$^{-2}$sec$^{-1}$sr$^{-1}$ is defined by $I_0(E_0)\equiv c\cdot n_\mathrm{particles}(t_0,E_0)/(4\pi)$ as in Ref.~\cite{Carr:2009jm}. When plugging in the solutions of the differential system~\eqref{Eq.Boltzmann}\footnote{The particular solution to the initial value problem of the form $\dot{n}(t)+3H(t)n(t)=S(t)$ for $t\ge t_1$ with $n(t_1)=0$ and source function $S(t)$ is
	\begin{align*}
		n(t)=\int_{t_1}^t \left[{1+z(t)\over1+z(t')}\right]^3S(t')dt',
	\end{align*}
	where $z=z(t)$ is the redshift-time relation. Usage of this result leads to the solution for the number density evolution of the species we need to evaluate for the photon flux.
	} \ we get the final formula for the photon flux:
	\begin{widetext}
	\begin{align}
		I_0(E_0)={c\over4\pi H_0^2}{\gamma_\mathrm{mrg}}E_0\int_{z_\mathrm{min}=0}^{z_\mathrm{begin}}{dz{\cal D}((1+z)E_0)\over E(z)}\int_z^{z_\mathrm{begin}}{dz'\exp\left[{t(z')-t(z)\over\tau_\mathrm{Hawk}}\right]\Theta(z'-z_\mathrm{end})\over E(z')(1+z')^4}
	\label{Eq.PhotonFlux}
	\end{align}
	\end{widetext}
	We have defined the auxiliary function $E(z)\equiv\sqrt{\Omega_V+(1+z)^3\Omega_{M}}$ \cite{Hogg:1999ad}, neglecting the radiation density, where $\Omega_V\simeq0.7$ and $\Omega_{M}\simeq0.3$ are the present energy densities of vacuum and matter relative to the critical density $\rho_{cr}$. Furthermore, the time difference that appears in the exponential in Eq.~\eqref{Eq.PhotonFlux} is calculated to be $t(z')-t(z)\simeq (0.8/H_0)\cdot {\rm ArcTanh}\left(1.2\sqrt{0.7+0.3(1+z)^3}\right) \Big|_{z}^{z'}$.
	At last but not least, we take into account the fact that there is a distribution of cluster sizes and there is a different number $N$ of PBHs that populate each cluster. We implement the halo mass function numerically numerically verified in Ref.~\cite{Inman:2019wvr} in the context of $\Lambda$CDM cosmologies.
	
\begin{figure}
		\centering
		\includegraphics[width=0.5\textwidth]{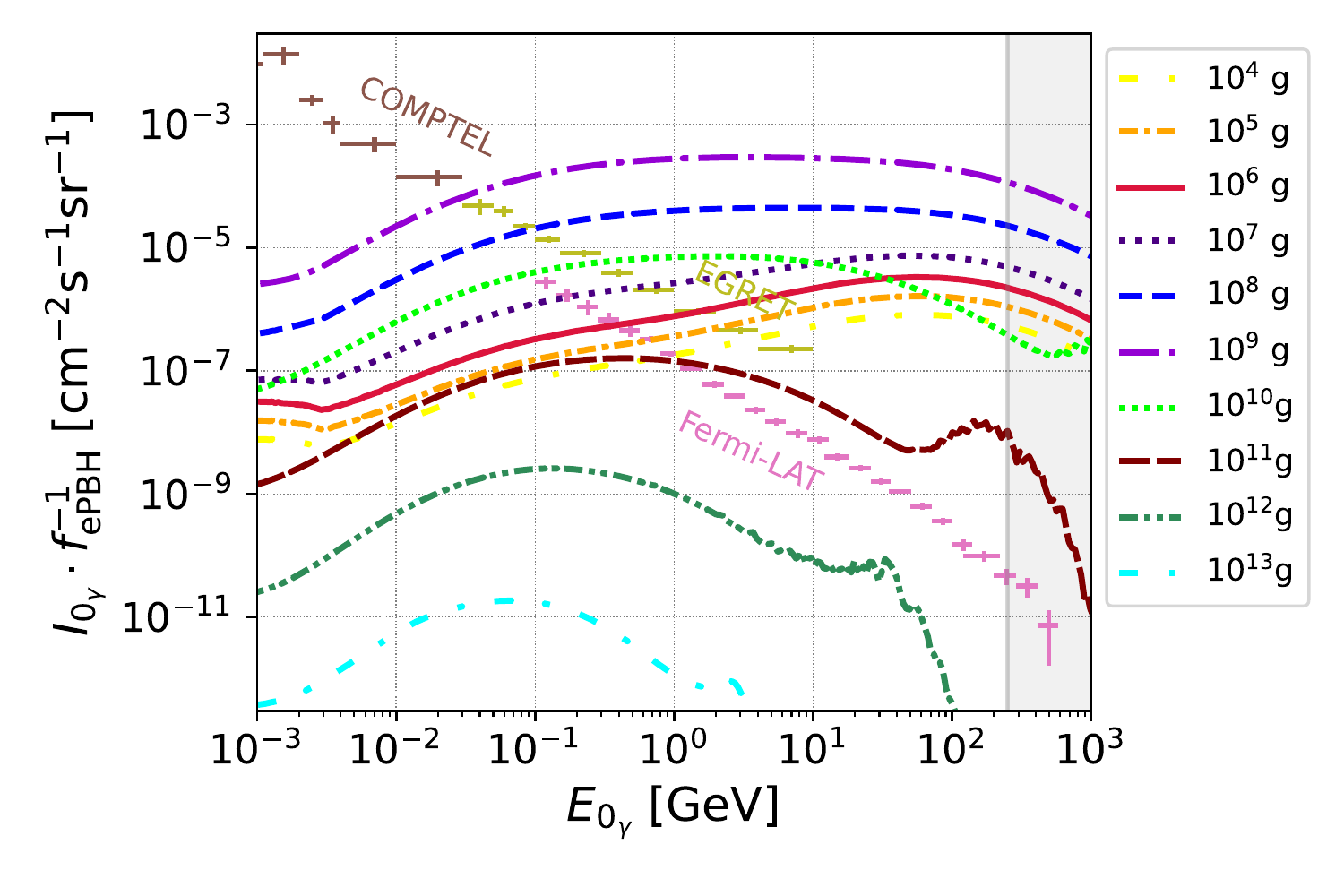}
		\caption{The unconstrained photon fluxes in units of cm$^{-2}$sec$^{-1}$sr$^{-1}$ normalized to the abundance of the PBHs, for ten values of the black hole mass scale (with color coding and linestyle as in the legend of the plot). See Table~\ref{tab:constraints} for the upper bounds on the $f_{\rm ePBH}$. All lines correspond to the total (secondary plus primary component) of the emission spectra. Also shown are the isotropic diffuse gamma ray emission spectra observed by Fermi-LAT (pink data points), EGRET (olive data points) and COMPTEL (brown data points). Also, the vertical dashed line indicates the high energy gamma ray attenuation point at $\sim250$GeV.}
		\label{Fig:photon_a}
	\end{figure}

	According to Ref.~\cite{1990PhRvD..41.3052M} (their Table~(VII)) the peak photon flux has an ${\cal D}\propto m^{-1.3}$ power law dependence on the mass of the evaporating black hole. In the previous section we have shown that the merger rate density is proportional to $m^{-3/5}$ (cf.~Eq.~\eqref{Eq.MergerRateDensity}). When these two effects are taken into account, the total photon flux has an $I_0\propto m^{-1.9}$ dependence. This overall effect can be seen in Fig.~\ref{Fig:photon_a} for $m>10^9\rm g$ and as the mass scale of the PBH population is reduced the photon flux increases and it peaks into larger and larger energies. However, for $m<10^9\rm g$ the flux starts to decline, because ${\cal D}$ has a positive slope dependence on mass in the frequency range of interest as the spectrum is expected to become harder and harder.
	
\begin{figure}
		\centering
		\includegraphics[width=0.5\textwidth]{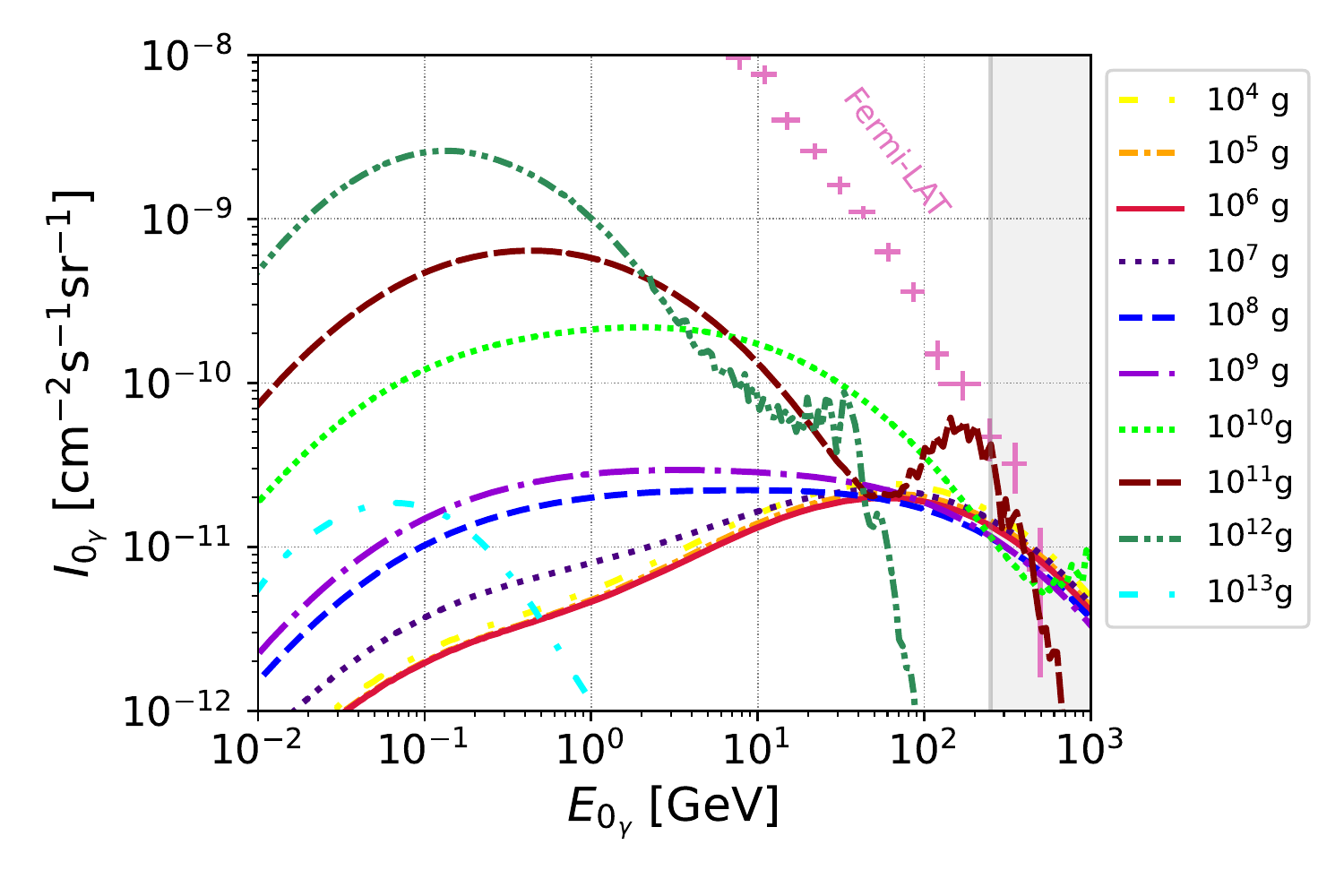}
		\caption{Same as Fig.~\ref{Fig:photon_a} but with the constrained taken into account. Bound from Fermi-LAT provides the most stringent upper limit on $f_{\rm ePBH}$.}
		\label{Fig:photon_b}
	\end{figure}

	In Figs.~\ref{Fig:photon_a} and~\ref{Fig:photon_b}, we show the resulting photon flux from evaporating remnants over cosmic time for ten mass scales of interest in the unconstrained ($f_{\rm ePBH}=1$) and constrained cases (that is, after placing the upper limit allowed on $f_{\rm ePBH}$).
	Furthermore, highly energetic photons with energies above $\sim250$GeV interact strongly with the background radiation field and the gamma rays get absorbed with an optical depth of $\simeq900$Mpc \cite{Kneiske:2003tx}. Because of this attenuation effect, the hard component gamma ray flux calculated
	should be considered for $E_0<250$GeV because flux is contributed by Hawking emission which takes place at high redshifts. However, this is not an issue for large-$N$ clusters because the highest flux comes from black hole remnants which form and evaporate at small redshifts. In any case, we constrain the gamma ray flux based on the bound set by Fermi-LAT in the high energy part of the spectrum.
	
\begin{figure}
		\centering
		\includegraphics[width=0.5\textwidth]{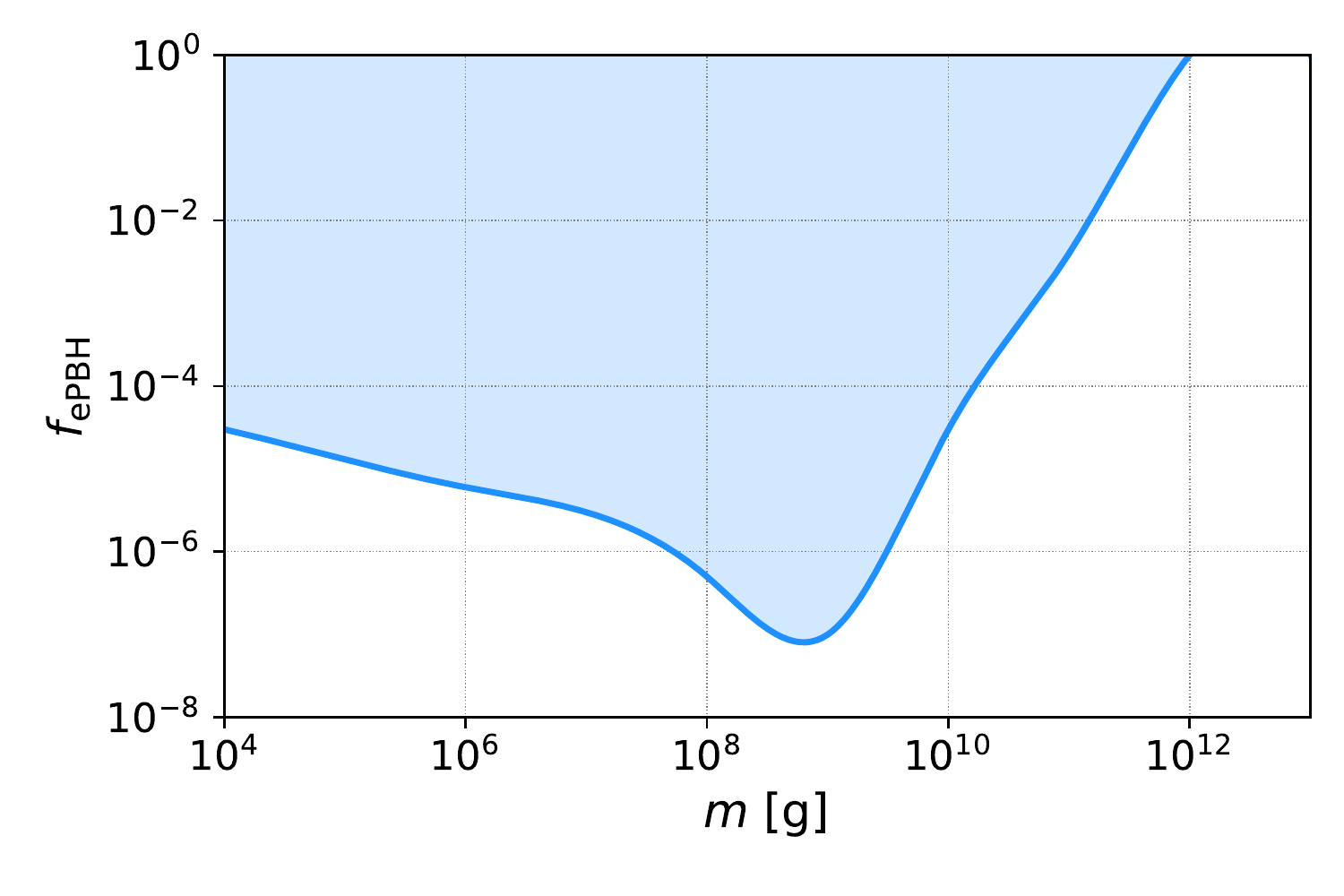}
		\caption{Constraint on the abundance of extramal PBHs, $f_{\rm ePBH}$. We interpolate the values of Table~\ref{tab:constraints} with a cubic fit. The shaded region above the curve is excluded by extragalactic gamma ray observations.}
		\label{fig.constraint}
	\end{figure}

	To respect the experimental bounds on the extragalactic diffuse gamma ray flux observed by gamma ray telescopes such as HEAO-1 \cite{Gruber:1999yr}, COMPTEL \cite{1996A&AS..120C.619K}, EGRET \cite{EGRET:1997qcq} and Fermi-LAT \cite{Fermi-LAT:2014ryh} we need to constrain the abundance of extremal black holes. Inspection of Figs.~\ref{Fig:photon_a} and~\ref{Fig:photon_b} provides an estimate for the largest possible value of the abundance of extremal black holes $f_\mathrm{ePBH}$ at mass scale $m$. In the mass range of interest which contributes with high fluxes, it is the secondary component of the emission which provides the most stringent upper bound. Moreover, Fermi-LAT places the tightest constraint because for light PBHs, the emitted spectrum becomes harder with decreasing mass. In Table~\ref{tab:constraints} we summarize the constraints on the abundance of these maximally charged black holes and give the upper bound $f_\mathrm{ePBH}^\mathrm{max}$. More specifically, if PBH clusters survive until the present epoch then the constraint can be evaded for $m>10^{13}\rm g$ and in principle it could be that $f_\mathrm{ePBH}=1$ for these masses. For lower mass scales, the maximum value of the abundance is constrained significantly and extremal PBHs can account for only a small fraction of the DM. On the other hand, if clusters tend to dissolve early  at higher redshifts, the constraint on $f_\mathrm{ePBH}^\mathrm{max}$ can be relaxed and for instance in the case of Fig.~\ref{Fig:photon_b} becomes stringent only for $m<10^{11}\rm g$. To place constraints on this case we only consider the piece of the flux below 250GeV due to attenuation effects at higher energies. We summarize the upper bound on the abundance of extremal PBHs in Fig.~\ref{fig.constraint}.
	
	\begin{table}[]
		\centering
		\begin{tabular}{c c}
			\hline
			$m$ [g] & $f_{\rm ePBH}^{\rm max}$\\
			\hline
			\hline
			$\ge10^{12}$ & 1 \\
			$10^{11}$    & $4\times10^{-3}$ \\
			$10^{10}$    & $3\times10^{-5}$ \\
			$10^{9}$     & $10^{-7}$ \\
			$10^{8}$     & $5\times10^{-7}$ \\
			$10^{7}$     & $3\times10^{-6}$ \\
			$10^{6}$     & $6\times10^{-6}$ \\
			$10^{5}$     & $1.3\times10^{-5}$ \\
			$10^{4}$     & $3\times10^{-5}$ \\
			\hline
		\end{tabular}
		\caption{Upper bound on the abundance of maximally charged PBHs $f_{\rm ePBH}$ relative to the DM for various mass scales $m$ for the PBHs.}
		\label{tab:constraints}
	\end{table}
	
	\subsection{Induced gravitational wave background}
	
	We have seen that observational signatures exist for light extremal PBHs. Constraints on the abundance of such black holes are placed and the fluxes induced from evaporating merger remnants could be detected if well-modeled contributions of other astrophysical origin are carefully subtracted \cite{Fornasa:2015qua,Blanco:2021icw}.
	However, the photon flux created from a population of extremal black holes heavier than about $10^{13}\rm g$ is estimated to be very low. This enables the possibility for accounting for the totality of dark matter. Nevertheless, gravitational waves are induced to second order by primordial curvature perturbations associated with the formation of the PBHs in the early Universe and the stochastic background they create might be observed by future gravitational wave detectors \cite{2010PhRvD..81b3517B}. These gravitational waves are an alternative signature associated with PBHs and could be detected by current or future gravitational wave experiments.
	
	The stochastic gravitational wave background is characterized in terms of its energy density per unit logarithmic frequency interval relative to the critical density $\rho_{cr}$ of the Universe, $\Omega_{gw}(f)=d(\rho_{gw}/\rho_{cr} )/d\ln f$ be it of astrophysical or cosmological origin \cite{Christensen:2018iqi}. To linear order, tensor and scalar metric perturbation modes are decoupled, but couple at higher orders \cite{Ananda:2006af}. This coupling leads to generation of gravitational waves induced at second order and by the nature of their generation are stochastic\footnote{For a recent review on scalar induced gravitational waves see Ref.~\cite{Domenech:2021ztg}.}. Probing the induced gravitational wave (IGW) background enables us to indirectly study the curvature perturbations responsible for PBH formation. The $\Omega_\mathrm{IGW}$ quantity for IGWs is related to the power spectrum of curvature perturbations, ${\cal P}_{\zeta}(k)$. A peak structure in ${\cal P}_{\zeta}(k)$ at $k=k_p$ translates into a peak in the IGW spectrum at $k\simeq k_p$ \cite{Saito:2008jc}. Nevertheless, we stress that the IGW spectrum is continuous even if it is sourced by a single mode in the scalar power spectrum. Ref.~\cite{Saito:2008jc} calculates the relation of the peak frequency of the second order IGWs to the mass of the primordial black hole given by\footnote{We have not included factors associated with the relativistic degrees of freedom and the parametrization of the deviation of PBH mass from horizon mass, as these would be order unity factors modifying the peak frequency of IGWs \cite{Nakama:2016gzw}.} $f_\mathrm{IGW}\sim10\mathrm{Hz}\cdot(m/10^{15}\rm g)^{-1/2}$. We remark that light PBHs with masses below $10^{13}\rm g$ form from curvature perturbations which induce gravitational waves that today would have high peak frequencies  far from the optimal sensitivity of current and proposed third-generation ground-based detectors. Thus, IGWs are well suited to probe for PBHs with masses above $10^{13}\rm g$.
	
	Semi-analytic formulas for the GW spectrum induced during the radiation-dominated era are provided in Ref.~\cite{Kohri:2018awv,Pi:2020otn} for the cases of monochromatic and log-normal power spectra of curvature perturbations. We utilize the fitting formula in Eq.~(29) from Ref.~\cite{Kohri:2018awv} for the IGW spectrum which agrees with numerical results. The spectrum of the IGWs at the present epoch is given by $\Omega_\mathrm{IGW,0}(f)=	1.6\times10^{-5}\cdot\Omega_\mathrm{IGW}(f)$ \cite{Fumagalli:2020nvq}.
	
	The fraction of the density of the Universe which collapses into extremal PBHs at the moment of their formation during the radiation dominated epoch is given by $\beta_\mathrm{ePBH}(k)=\mathrm{erfc}(0.1507/\sqrt{{\cal P}_\zeta(k=aH)})$ where the power spectrum of curvature perturbations ${\cal P}_\zeta$ is evaluated at horizon crossing (cf.~(6.5) from Ref.~\cite{Harada:2013epa}) and where erfc is the complementary error function. This formula for $\beta_\mathrm{ePBH}$ assumes that the density perturbations follow the Gaussian probability distribution. Therefore, it is evident that the largest abundance comes from those scales for which the power spectrum becomes maximum at the peak wavenumber $k_p$. Taking a monochromatic\footnote{The monochromatic function peaked at $k=k_p$ is parametrized as follows
	\begin{align*}
		{\cal P}_\zeta(k)={A_\zeta}\delta\left[{\ln^2(k/k_p)}\right],
	\end{align*}
	where $A_\zeta$ is the dimensionless amplitude and $\delta(\_)$ is the Dirac delta.} shape for ${\cal P}_\zeta$, then to a good approximation $\beta_\mathrm{ePBH}\simeq\mathrm{erfc}(0.1507/\sqrt{A_\zeta})$. The value of $\beta$ can be related to the abundance of PBHs in the late Universe at the present epoch $f_\mathrm{ePBH}$ via $\beta_\mathrm{PBH}\simeq (a_\mathrm{form}/a_{eq})f_\mathrm{PBH}\simeq(m/M_{eq})^{1/2}f_\mathrm{PBH}$ where $a_{eq}$ is the scale factor at matter-equality and $a_\mathrm{form}$ at the formation epoch of the PBHs. This relation follows from the realization that after  matter-radiation equality, the abundance of PBHs does not vary relative to the total energy density since it is dominated by matter. The mass at equality is $M_{eq}\simeq2.8\times10^{17}M_\odot$ \cite{Nakama:2016gzw}. Therefore, in order to achieve $f_\mathrm{PBH}=1$, the abundance of the PBHs at the era of their formation should have been $\beta_\mathrm{ePBH}\simeq4.1\times10^{-21}\cdot(m/10^{10}\rm g)^{1/2}$. Finally, the amplitude of the power spectrum of curvature perturbations becomes
	\begin{align}
		A_\zeta\simeq{ 0.023\over\left\{\mathrm{erfc}^{-1}\left[4.1\times10^{-21}\left({m\over10^{10}\rm g}\right)^{1/2}\right]\right\}^2 }
		\label{Eq.Azeta}
	\end{align}
	with $f_\mathrm{ePBH}=1$.
	The function erfc$^{-1}$ is the inverse of the complementary error function.
	
\begin{figure}
		\centering
		\includegraphics[width=0.5\textwidth]{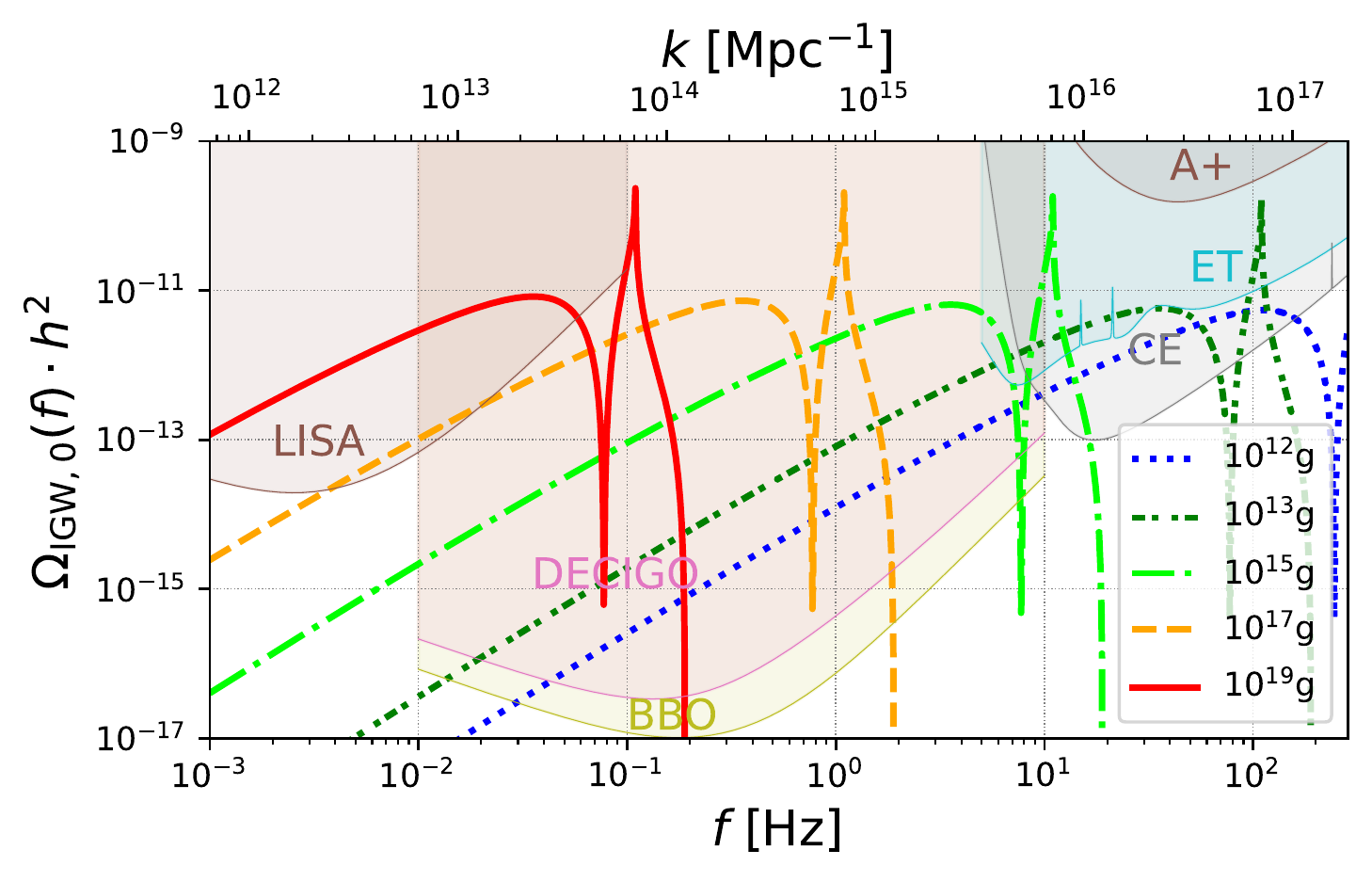}
		\caption{The second order scalar-induced gravitational wave stochastic background assuming specific values of the PBH mass as a function of frequency. We also include a second x-axis for the wavenumber. 
		Also shown are the sensitivity curves of a few future proposed/planned gravitational wave detectors \cite{Thrane:2013oya,Moore:2014lga,LIGOScientific:2016wof,Schmitz:2020syl}, along with the A+ design curve \href{https://dcc.ligo.org/LIGO-T1800042/public}{https://dcc.ligo.org/LIGO-T1800042/public}.}
		\label{Fig.IGWs}
	\end{figure}

	In Fig.~\ref{Fig.IGWs} we present our results for the monochromatic spectrum of curvature perturbations. One feature of the shape of IGWs is the $\propto f^{2}$ dependence on the infrared part of the spectrum which could be used to discriminate spectra from other stochastic gravitational wave backgrounds associated with different physical processes. This power law dependence is associated with the choice of the log-normal form of the power spectrum of curvature perturbations.
	Another feature of the spectra which stems from the formula in Eq.~\eqref{Eq.Azeta} is that it does not depend strongly on the mass scale of the PBH. In particular, in the mass range $10^{10}\rm g$ to $10^{20}\rm g$ the relative variation in the value of $A_\zeta$ is at most on the order of $30\%$. This is the reason why in Fig.~\ref{Fig.IGWs} the amplitude of the IGW spectra which is proportional to $A_\zeta^2$ is almost the same for each value of $m$.
	
	There are various gravitational wave experiments and detectors, some of which are operational and others proposed and/or planned to probe stochastic backgrounds among other resolvable sources. In the LIGO-Virgo band, for high frequencies in the range $\sim10-1 \rm kHz$ there is the ground-based network of LIGO \cite{LIGOScientific:2014pky}, Virgo \cite{VIRGO:2014yos} and KAGRA \cite{Aso:2013eba} as well as the proposed third generation Cosmic Explorer (CE) \cite{LIGOScientific:2016wof} and Einstein Telescope (ET) \cite{Sathyaprakash:2012jk} detectors. In the deci-Hz band, frequencies in the range $\sim10^{-1}-10\rm Hz$, the Decihertz Gravitational wave Observatory (DECIGO) \cite{2017JPhCS.840a2010S} and Big Bang Observer (BBO) \cite{Harry:2006fi} space-based detectors as well as atomic interferometer observatories such as AEDGE \cite{AEDGE:2019nxb}, MAGIS \cite{Graham:2017pmn} and AION \cite{Badurina:2019hst} have been proposed. In this interval, the IGWs associate with PBHs of mass scale $\sim10^{17}\rm g$. In the mHz frequency range the Laser Interferometer Space Antenna (LISA) \cite{2017arXiv170200786A}, TianQin \cite{TianQin:2015yph} and Taiji \cite{Ruan:2018tsw} are planned/proposed missions. These space-borne observatories would probe the stochastic IGW background associated with the formation of PBHs with masses of the order of $10^{20}\rm g$. In Fig.~\ref{Fig.IGWs} we also include the sensitivity curves of LISA, DECIGO, BBO, ET, CE and A+ which is the upgraded version of Advanced LIGO \cite{McCuller:2020yhw}.
	
	\section{Conclusions}
	\label{Sec:conclusions}
	
	To summarize, in this work we have studied the possibility that a population of light extremal black holes can comprise the dark matter as well as a few prospective signatures associated with them. Constraints on this model were placed from the Hawking evaporation of the non-extremal merger remnants and the gamma ray flux was calculated.
	
	We have demonstrated that if  clusters of near-extremal holes contain a sufficient number of members to survive up to low redshift, the
	hard photons from continued evaporation begin to  dominate the high energy  diffuse background. We find that the diffuse  photon flux can be observed for holes lighter than about $3\times10^{11}\rm g$. We note that the conservative bound was placed considering that the formation of late-type ePBH-ePBH binaries as the gamma ray flux from early mergers would be redshifted to undetectable levels.
	
	Such a flux might be detected by current gamma ray telescopes like Fermi-LAT and the atmospheric Cherenkov telescopes such as the HAWC \cite{HAWC:2017udy}, VERITAS \cite{Weekes:2001pd} and H.E.S.S. \cite{HESS:2005hpr} collaborations which can probe high energy gamma rays. For simplicity we have focused on a monochromatic mass spectrum at mass $m$ and probed the phenomenological consequences the clustering effect would have on the formation of PBH clusters after the epoch of recombination. We demonstrated that the electromagnetic bremsstrahlung is a more efficient process via which a binary of oppositely charged black holes is assembled inside such a PBH cluster compared to gravitational captures. As a consequence the merger rate is enhanced by a factor of $\sim250$ (see Eq.~\eqref{Eq:emVSgw}). We found that a model with extremal PBHs of mass smaller than $\simeq10^{13}\rm g$ is significantly constrained from the extragalactic diffuse gamma ray background measured by various gamma ray telescopes if PBH clusters survive until today. This constraint could be relaxed if the clusters of PBHs had a smaller lifetime (see Fig.~\ref{Fig:photon_b} for an example with clusters of $N=200$ members). For instance in that case the model is constrained only for $m<10^{11}\rm g$. After clusters dissolve, the extremal PBHs would then be released in the field and the merger rate from dynamically assembled pairs would drop. The gamma ray flux produced by extremal PBHs with masses larger than $\simeq10^{13}\rm g$ is not large and does not come in conflict with the observations. To probe these masses, the stochastic gravitational wave background induced by primordial curvature perturbations at the epoch of PBH is an alternative signal and could be detected by future proposed and planned gravitational wave observatories like the space-born DECIGO, BBO and atomic interferometers for $m=10^{15}-10^{20}\rm g$ and the ground-based CE and ET for $m=10^{13}-10^{15}\rm g$. The detection of a multi-messenger signal from such a population of charged black holes would be of high significance and the properties of initial conditions of the early Universe and the nature of DM would be probed.
	
	\begin{acknowledgments}
	
    We would like to thank the anonymous referee for  comments. We are also grateful to E.~Berti for reviewing and providing feedback that improved the manuscript and M.~Kamionkowski for kindly reading the first version of the draft. Furthermore, we thank T.~Helfer, M.~Cheung and L.~Reali for their useful discussions.
	
	\end{acknowledgments}
	
	\appendix
	
	\section{Lifetime of black hole plasmas}
	\label{App:relax}
	
	To calculate the two-body relaxation of such clusters with oppositely charged black holes, we may apply a theory similar to the one used to find corresponding formulae in the context of clusters with neutral objects. The novel feature here would just be the inclusion of the electromagnetic force involved during two-body encounters.
	
	In the following we proceed to calculate the relaxation timescale in the context of extremal black hole clusters. Following closely the discussion in Ch.~1 of \cite{2008gady.book.....B}, the perpendicular force acted on a subject black hole by a field black hole traveling at a relative velocity $v$ and encountering the subject hole at an impact parameter of $b$ would be $F_\perp=(\alpha/b^2)[1+(vt/b)^2]^{-3/2}$. This force may be repulsive or attractive depending on the signs of the charges. Ideally, during a two-body encounter, time $t$ runs from $-\infty$ to $+\infty$ and the point of closest approach would correspond to the $t=0$ time. The assumption made here, would be that the trajectory of the reduced mass is not deviating from that of a straight line; in other words, the interaction is not strong and the deflection can be neglected. We remind the reader that the parameter $\alpha$ is defined to be $Gm_1m_2-KQ_1Q_2$ as a combination of the masses and charges of the interacting objects. On average, since the mean charge per unit volume is zero, a black hole of a given charge would encounter an equal amount of positive and negative charges per unit time. The magnitude of the velocity change is then found by integrating the impulse over time and dividing it by the reduced mass of the system $m/2$ and is given by $\delta v=4\alpha/(bmv)$. If $\sigma=v/\sqrt{2}$ is the typical velocity dispersion in the cluster, and $R$ is the radius of that cluster, then $\tau_\mathrm{cross}=R/\sigma$ would be the crossing time. The mean-square velocity change per crossing time is $\Delta v ^2=32N\alpha^2\ln\Lambda/(mvR)^2$, where $\ln\Lambda$ is the Coulomb logarithm. The Coulomb logarithm would be given to a good approximation by $\ln\Lambda\simeq\ln(R/b_{90})$ where $b_{90}=4\alpha/(mv^2)$ is the impact parameter for which the deflection of the field hole would be large on the order of $90^{o}$ either away or towards the subject hole. We define the relaxation timescale as follows (as defined in \cite{2008gady.book.....B}):
	\begin{align}
		\tau_\mathrm{relax}\equiv{v^2\over\Delta v^2}{\sqrt{2}R\over v}\simeq{0.05N\over\ln N}\tau_\mathrm{cross}
		\label{Eq.RelaxationTimescale}
	\end{align}
	where we have used $\sigma^2=(6/5)(GNm/R)$ the Virial theorem (see footnote 1).
	This would be the timescale after which the initial conditions of a subject black hole is lost via the cumulative effect of two-body encounters perturbing the velocity. The lifetime of the cluster would correspond to the timescale after which it completely dissolves as it slowly evaporates due to the high velocity tail of the Maxwell-Boltzmann distribution and it is given by $\tau_\mathrm{life}\simeq140\tau_\mathrm{relax}\simeq6.3\cdot\tau_\mathrm{cross}N/\ln N$ \cite{Afshordi:2003zb}. Therefore, clusters of charged black holes evolve faster and have a smaller lifetime than the corresponding clusters of neutral objects, by a factor of about 2. Using the cluster parameters from Ref.~\cite{Jedamzik:2020ypm} the lifetime of the cluster then becomes
	\begin{align}
		\tau_\mathrm{life}\simeq0.15\mathrm{Myr}\cdot{N^{7/4}\over\ln N},
		\label{Eq:ClusterLife}
	\end{align}
	which only depends on the number of the members in the cluster, $N$. The cluster survives for longer than a Hubble time if it contains at least $N\simeq2000$ members.
	
	\section{Electromagnetic bremsstrahlung}
	\label{App:bremss}
	
	In this Appendix we calculate the total electromagnetic energy radiated during the close encounter of two charged black holes. The calculation we perform below bears similarity with the one followed in \cite{1977ApJ...216..610T}.
	
	Consider two point charged particles with masses $m_i$, electric charges $Q_i$ and position vectors $\boldsymbol{r}_i$ respectively, with $i=1,2$. The symbol $m_{12}$ denotes the total mass of the two-body system $m_1+m_2$. We work in the center of mass system so that $m_1\boldsymbol{r}_1+m_2\boldsymbol{r}_2=\boldsymbol{0}$ and consider the relative motion of the reduced mass $\mu=m_1m_2/m_{12}$. We also define the symbol $\boldsymbol{r}=\boldsymbol{r}_1-\boldsymbol{r}_2$ to denote the relative position of the two bodies. The electric dipole of the two particles is then written as $\boldsymbol{d}=Q_1\boldsymbol{r}_1+Q_2\boldsymbol{r}_2=q\boldsymbol{r}$, in which $q\equiv(Q_1m_2-Q_2m_1)/m_{12}$ is the effective charge of the system. The two particles are assumed to move initially on unbound orbits. During the two body approach we neglect the influence of third bodies which is unimportant when the interaction is close and occurs on a timescale much shorter than two-body relaxation. 
	
	Two useful integrals of the motion of the system are the total energy $\cal E$ and angular momentum $\cal J$. 
	The total energy of the two-body system is parametrized in terms of the orbital eccentricity $e$ and the pericenter $r_p$ of the encounter as follows
	\begin{align}
		{\cal E} = {\alpha\over2r_p}(e-1)
	\label{Eq.PairEnergy}
	\end{align}
	where we have defined the parameter $\alpha\equiv GM\mu-KQ_1Q_2$, the symbol $G$ is the Newton gravitational constant and $K$ is the Coulomb force constant. For future reference, the symbol $c$ is devoted for the speed of light. The angular momentum is correspondingly given by
	\begin{align}
		{\cal J} = \Big(\alpha\mu r_p(1+e)\Big)^{1/2}.
	\end{align}
	It becomes evident that if the two interacting particles have opposite electric charges, the encounter as an unbound system is well defined. However, when they are like charges, the theory would still work as long as $\alpha\ge0$. In the case of extremal black holes of like charges, the charge $Q_i$ and mass $m_i$ of each black hole are related via $\sqrt{K}Q_i=\pm\sqrt{G}m_i$ and the system has zero total energy and angular momentum. The black holes then necessarily approach along a parabolic orbit. On the other hand, if the two extremal black holes have opposite charges then the parameter $\alpha$ obtains its maximum value of $2Gm_1m_2$.
	
	In the following analysis, we approximate the hyperbolic encounter by a parabolic one. This approximation is similar to the one made in \cite{1989ApJ...343..725Q}. The reasoning behind this one is the fact that the relative acceleration maximizes when the two particles are at the closest approach point. In this regime, when $r\sim r_p$, the electromagnetic and gravitational radiation becomes the largest and a hyperbola can be approximated by a parabola. For this reason we set the orbital eccentricity equal to unity, $e=1$. This greatly simplifies our analysis when we calculate the total energy radiated in electromagnetic waves during the encounter. Furthermore, the equations of the relative motion is determined by the following equations relating the relative position $r$ and true anomaly $\phi$,
	\begin{subequations}
		\begin{align}
			&r={2r_p\over1+\cos(\phi)}\label{Eq:motion_r},\\
			&\mu r^2\dot{\phi}={\cal J}\label{Eq:motion_phi}.
		\end{align}
	\end{subequations}
	The overdot in the expression above denotes time derivative.
	
	The amount of gravitational wave energy emitted along an unbound orbit was calculated in \cite{1977ApJ...216..610T}. Under the approximation of parabolic approach and with a mild modification accounting for the charge of the black holes we have
	\begin{align}
		\delta E_{gw} = {85\pi\over12\sqrt{2}}{G\over c^5}{\alpha^{5/2}\over \mu^{1/2} r_p^{7/2}}.
		\label{Eq:EnergyGW}
	\end{align}
	
	To obtain the electromagnetic energy released during the unbound encounter, we need to integrate the electromagnetic power loss over the orbit of the reduced body. The total power is given by the relativistic generalization of Larmor's formula and assuming small black hole velocities relative to the speed of light, $v=\|\boldsymbol{v}\|\ll c$, we approximate this by the dipole radiation term $(2/3)K\ddot{d}^2/c^3$,
	\begin{align}
		P_{em}&={2\over 3}{Kq^2\over c^3}\left(1-{v^2\over c^2}\right)^{-3}\left\{ \|\dot{\boldsymbol{v}}\|^2 - {1\over c^2}\|\boldsymbol{v}\times\dot{\boldsymbol{v}}\|^2 \right\}\nonumber\\&\simeq{2\over 3}{Kq^2\over c^3}\ddot{r}^2.
		\label{Eq:EmPower}
	\end{align}
	The total energy radiated is obtained by integrating the Eq.~\eqref{Eq:EmPower} over all time, or over the true anomaly when changing the variable of integration,
	\begin{align}
		E_{em} = \int_{-\infty}^{+\infty}P_{em}dt = \int_{-\pi}^{+\pi}P_{em}{d\phi\over\dot{\phi}}.
	\end{align}
	Utilizing Eq.~\eqref{Eq:motion_r} and~\eqref{Eq:motion_phi} the detailed calculation of the previous integral yields the final formula for the electromagnetic energy dissipated away,
	\begin{align}
		\delta E_{em} \simeq {7\pi\over24\sqrt{2}}{Kq^2\alpha^{3/2}\over c^3\mu^{3/2}}{1\over r_p^{5/2}}.
		\label{Eq:EnergyEM}
	\end{align}
	
	\section{Properties of captured pairs}
	\label{App:CaptureMeansMerger}
	
	The angular momentum lost due to electromagnetic bremsstrahlung is $\dot{\boldsymbol{J}}_{em}=-(2Kq^2/(3c^3))(\dot{\boldsymbol{r}}\times\ddot{\boldsymbol{r}})$. The total variation of the angular momentum $\delta J$ is of the order of $v^2/c^3$. The initial angular momentum of the two-body system is $J=b\mu v_\mathrm{rel}$ where the square of the impact parameter would be $b^2=2r_p\alpha/(\mu v_\mathrm{rel}^2)$ in the focusing regime. Thus, $\delta J/J\sim O(v/c)^3$ and therefore, we may neglect $\delta J$ relative to $J$ itself altogether. Following the closely related discussion considered in \cite{OLeary:2008myb} which treats gravitational bremsstrahlung, it is a good approximation to analogously set $|{\cal E}_\mathrm{fin}|
	\simeq \delta E_{em}$ for electromagnetic captures. Combining this approximation with the fact that the final energy after the close encounter becomes ${\cal E}_\mathrm{fin}=(1/2)m\sigma^2-\delta E_{em}<0$ the semimajor axis $a_0$ and eccentricity $e_0$ of the newly-formed pair of oppositely charged extremal black holes will be given by the following two expressions:
	\begin{subequations}
		\begin{align}
			&a_0 = {\alpha\over2|{\cal E}_\text{fin}|} = {3\sqrt{2}\over7\pi}r_p\left({r_p\over r_H}\right)^{3/2}<2.1\cdot\left({c\over\sigma}\right)^2r_H,\label{Eq.SmaNew}\\
			&e_0 = \sqrt{1-{2|{\cal E}_\text{fin}|J^2\over \mu\alpha^2}}\nonumber\\&\Leftrightarrow1-e_0^2= {14\pi\over3\sqrt{2}}\left({r_H\over r_p}\right)^{3/2}> 2.5\cdot\left({\sigma\over c}\right)^{6/5}  \label{Eq.EccenNew}.
		\end{align}
	\end{subequations}
	The inequalities come from setting the pericenter to its maximum allowed value for capture to occur in the em regime (since electromagnetic captures are dominating).
	We observe that since the pericenter for capture is a large multiple of the horizon size of the black holes (cf.~Eq.~\eqref{Eq:pericenter}), the eccentricity of a captured pair is expected to be very close to unity. Furthermore, it is exceptionally rare for two objects to interact with a very small pericenter value, therefore the semimajor axis tends to be closer to its maximum.
	
	According to Ref.~\cite{Liu:2020cds} the merger timescale of a binary composed of charged black holes in the high eccentricity approximation, $e_0\simeq1$, is given by
	\begin{align}
		\tau_\text{mrg}=\min\left\{ {3\over85}{c^5a_0^4(1-e_0^2)^{7/2}\mu\over G\alpha^2} , {c^3a_0^3(1-e_0^2)^{5/2}\mu^2\over K\alpha q^2} \right\}.
		\label{Eq.mergerTime}
	\end{align}
	In the near-extremal case, if the charges of the black holes in the binary have the same sign, then the evolution is prolonged and the timescale becomes even larger than the binary-single interaction time. Then, the binary tends to ionize as a consequence of these encounters because it is marginally soft\footnote{A binary in a collisional environment is called soft if the kinetic energy of an ambient single object exceeds the binding energy of the binary. Otherwise it is called hard. Soft binaries tend to loosen with time and hard binaries tighten \cite{1975MNRAS.173..729H}.}. In particular, the hard-soft boundary is at $\sim m\sigma^2$ and the binding energy of the captured pair would be $\sim m\sigma^2$ as well to within $O(1)$ prefactors.
	
	The merger time-scale in Eq.~\eqref{Eq.mergerTime} has to be compared with the typical timescale for interaction of the newly formed binary with a third black hole. Such a binary-single encounter would perturb the structural parameters of the binary and could lead to its ionization, reducing significantly the merger rate \cite{Jedamzik:2020ypm}. The binary-single interaction timescale would be given by $\tau_\mathrm{inter}=(n\langle\Sigma_\mathrm{inter} v_\mathrm{rel}\rangle)^{-1}$ which when evaluated with typical cluster parameters gives
	\begin{align}
		\tau_\mathrm{inter}\simeq0.12\cdot\left({R\over a_0}\right)\tau_\mathrm{cross}>0.06\cdot\left({R\over r_H}\right)\left({\sigma\over c}\right)^2\tau_\mathrm{cross}.
		\label{Eq.EncTimescale}
	\end{align}
	The lower bound arises because we have considered a binary with the largest semimajor axis is could have formed with via electromagnetic bremsstrahlung.
	Above in Eq.~\eqref{Eq.EncTimescale} we have only included the focusing term for the interaction cross section $\Sigma_\mathrm{inter}$, because the encounter is assumed to be strong and for that we used the binary-single pericenter distance to be on the order of the binary's semimajor axis.
	
	From Eq.~\eqref{Eq.mergerTime} in the em regime we have $\tau_\mathrm{mrg}<11\cdot(r_H/c)(c/\sigma)^3$. When compared with the mean binary-single timescale from Eq.~\eqref{Eq.EncTimescale} we obtain that $\tau_\mathrm{mrg}/\tau_\mathrm{inter}<(12/N)^2$.
	Thus we have demonstrated that electromagnetic captures lead to a merger on a timescale much shorter than the binary-single interaction timescale required to perturb the pair given that the cluster contains more than a dozen members. Extremal black holes which capture themselves electromagnetically may be wide, however their eccentricity is so close to unity and electromagnetic dipole emission so efficient that the binaries are thought to merge quickly upon capture. For this reason, the calculation of the capture rate provides an estimate for the merger rate, since every captured pair essentially promptly merges.
	
	\bibliography{refs}
	
\end{document}